\documentclass[aps,prl,twocolumn]{revtex4}

\usepackage{latexsym,amssymb,float}

\usepackage{graphicx}
\usepackage{dcolumn}% Align table columns on decimal point
\usepackage{bm}% bold math

\begin{document}

\title{Magnetic field dependence of the
superconducting gap node topology
in non-centrosymmetric CePt$_3$Si}

\author{I. Eremin}

\affiliation{ Max-Planck Institut f\"{u}r Physik Komplexer Systeme,
D-01187, Dresden, Germany. \\
 Institut f\"{u}r Mathematische/Theoretische Physik, Technische Universit\"{a}t
Carlo-Wilhelmina zu Braunschweig, 38106 Braunschweig, Germany.
}

\author{J.F. Annett}

\affiliation{ H.H. Wills Physics Laboratory, University of Bristol, Tyndall
Avenue, Bristol, BS8 1TL, UK }

\begin{abstract}
The non-centrosymmetric superconductor CePt$_3$Si is
believed to have a line node in the energy gap
arising from coexistence of s-wave and
p-wave pairing.
We show that a weak c-axis magnetic field will remove this line
node, since it has no topological stability
against time-reversal symmetry breaking perturbations.  Conversely a field in the
$a-b$ plane is shown to remove the line node on some regions of the Fermi surface, while
bifurcating the line node in other directions, resulting in  two 'boomerang'-like
shapes. These line node topological changes  are
predicted to be observable experimentally in the low temperature heat capacity.
\end{abstract}

%\section{Introduction}

\maketitle

%\section{Introduction}

The existence of nodal points or lines in the energy gap of a
superconductor or a Fermi superfliud, such as $^3$He, is one of the
main characteristics of unconventional symmetry pairing
states.
An important
question is whether or not these nodes are accidental or are fully
required by the fundamental symmetry of the pairing state.  Secondly
one can ask whether or not the nodal points or lines are
topologically stable, or whether they can in principle be destroyed
by small perturbations. This question was discussed in a seminal
paper by Volovik\cite{Volovik1987}, in which he showed that point
nodes, such as in $^3$He-A, are diabolical points characterized by a
Berry phase and topological charge. Because of this property the nodal
structure cannot be removed by small perturbations, such as a
magnetic field. Very recently Sato\cite{Sato2006} and Volovik\cite{Volovik2006} have
independently,
shown that
for line nodes of the gap function topological stability is only guaranteed
if time reversal symmetry is preserved.
The topological structure of the nodal line can be classified as
$Z_2$ (i.e. the group, \{0,1\}, of integers modulo 2). Perturbations
breaking time reversal symmetry remove conservation of the relevant quantum
numbers, and hence line nodes are not topologically stable against such
perturbations.

It is especially interesting to consider the gap nodal structure in
the recently discovered non-centrosymmetric heavy fermion
superconductor CePt$_3$Si\cite{Bauer2004}. This material is found to
have a spin-density wave Neel order at $T_N=2.2$K, corresponding to
an effective moment $0.16\mu_B/Ce$\cite{Metoki2004} and then
superconductivity below about $0.75$K\cite{Bauer2004}. The
superconductivity is expected to be nearly unique because this
material is tetragonal with space group $P4mm$, having no center of
inversion symmetry. The existence of such an inversion center is
normally essential in the distinction between spin singlet and spin
triplet superconductivity in systems where the spin-orbit coupling
is
large\cite{Anderson1984,Annett1990,Sigrist1991,Capelle1999a}.
Therefore, without an inversion center this system can be expected
to possess simultaneous singlet and triplet pairing components on
the sheets of its, non Kramers degenerate, Fermi surface. Thermal
conductivity experiments suggest that the superconducting gap most
probably has lines of nodes on the Fermi surface\cite{Izawa2005},
while NMR experiments\cite{Yogi2004,Yogi2006} show evidence for
crystal field excitations, as well as the formation of a heavy
fermion metallic state. Surprisingly,
$1/T_1T$ does not show the usual $T^3$ behaviour of other Ce based
heavy fermion superconductors, but rather a weak Hebel-Slichter like
peak below $T_c$.

\begin{figure}[b]
\centerline{\scalebox{0.32}{\includegraphics{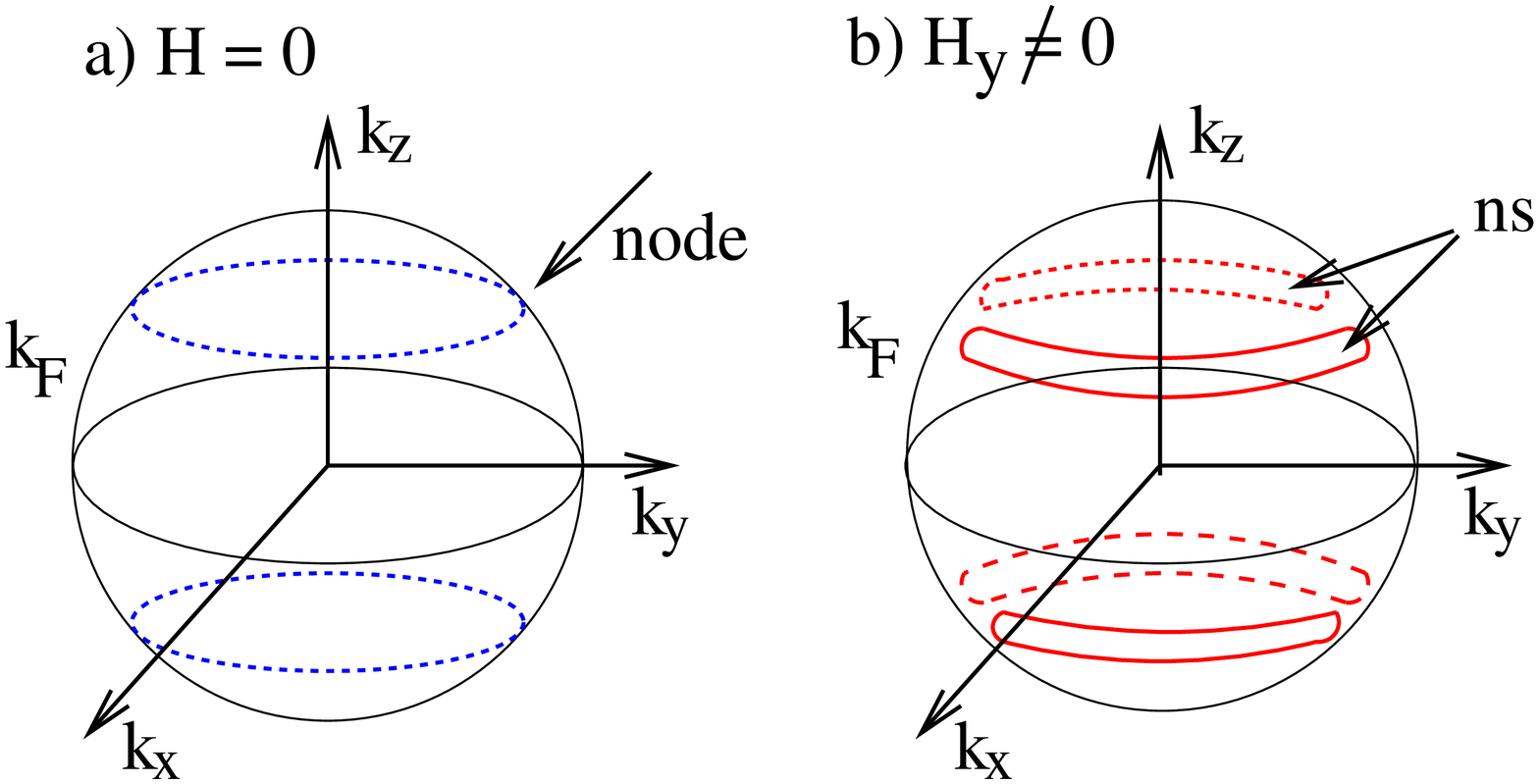}}}  %,width=8.5cm}
\caption{(color online) Illustration of the influence of the $ab$-plane external
magnetic field on the gap line nodes in
CePt$_3$Si.
%Note for the sake of simplicity we assume {\bf g}-vector
%is aligned along the $x$-direction.
 (a) zero-field line nodes for $\pm k_z = const$
on one Rashba Fermi surface sheet; (b) $H_y \neq
0$ removes the node along the $\pm k_y$ direction and bifurcates the node
elsewhere, yielding  ``boomerang''-like nodal topology. }
\label{fig1}
\end{figure}

In this letter we consider the effect time reversal symmetry breaking
on the nodal topology in  CePt$_3$Si. Starting with
the model gap  nodal structure proposed
 by Hayashi {\it et al.}\cite{Hayashi2006},
 we show
that  the line node can be completely removed by an arbitrarily weak  c-axis magnetic field.
In contrast, for $a-b$ plane oriented magnetic fields we show that
 magnetic fields create
additional lines of nodes, splitting the single degenerate node of
the time-reversal symmetric ground state into two. Moreover, we show
that the double line of nodes do not extend around the whole Fermi
sphere but form a topological ``boomerang''-like structure, as shown
in Fig.~\ref{fig1}. These effects lead to a dramatic increase in the
linear $T$ coefficient in $C_V/T$, which should  be visible
experimentally. The appearance of these extra line nodes in a
magnetic field and their large effect on the physical properties is
surprising given the often stated argument that the non Kramers
degenerate Fermi surface should suppress the effect of Zeeman
interactions on the superconducting state\cite{Frigeri2004b}.
However, a strong and unusual angle dependence to the paramagnetic
susceptibility has already been predicted by
Samokhin\cite{Samokhin2005} and Mineev\cite{Mineev2005} has shown
that the paramagnetic limit is expected to be strongly field-angle
dependent.
% Our work extends the $Z_2$ symmetry topological arguments
%given by Sato\cite{Sato2006} for CePt$_3$Si, by showing both node
%removal and node doubling depending on applied the field
%orientation.

The superconducting state of CePt$_3$Si has already been
subjected to considerable experimental and theoretical effort.
As well as the existence of the gap line node\cite{Izawa2005},
a surprisingly high $H_{c2}$ has been observed of about
 $3.2$T
for $H$ along (001) and $2.7$T for $H$ along (100)\cite{Yasuda2004},
exceeding the Pauli limiting field of $H_P \sim 1.2$T. In addition
to the weak antiferromagnetism, with $Q=(001/2)$ and an effective
moment of about $0.16\mu_B$ per Ce atom, considerably less than the
$2.54\mu_B/Ce$ expected for the Ce$^{3+}$ ion $4f_{5/2}$ ground
state, well defined crystal field excitations at $1$meV and $24$meV
were also seen\cite{Metoki2004}.

Theoretical models of the superconducting state
in CePt$_3$Si are based upon the existence of
a Rashba type spin-orbit coupling
\begin{equation}
  \hat{H}_{so} = \alpha \sum_{{\bf k},s,s'}
 {\bf g}_{\bf k}.\mathbf{\sigma}_{ss'} c^\dagger_{{\bf k}s}c_{{\bf k}s'}
\end{equation}
as studied by Gor'kov and Rashba\cite{Gorkov2001}. Here ${\bf
g}_{\bf k}= -{\bf g}_{-{\bf k}}$ is a real pseudovector, by
convention normalized so that $\langle g^2_{\bf k} \rangle_F = 1$
where $\langle \dots \rangle_F$ denotes an average over the Fermi
surface, $\mathbf{\sigma}=(\sigma_x,\sigma_y,\sigma_z) $ is the
vector of Pauli matrices and $c^\dagger_{{\bf k}s}$ is the usual
Fermi field operator. % for Bloch state ${\bf k}$ and spin $s$.
This
spin-orbit coupling removes the usual Kramers degeneracy between the
two spin states at a given ${\bf k}$, and leads to a quasi-particle
dispersion $\varepsilon_{\bf k}= \epsilon_{\bf k}\pm \alpha |{\bf
g}_{\bf k}|$ \cite{Frigeri2004},  splitting the Fermi surface into two
sheets with a ${\bf k}$ dependent local spin orientation. The
fully-relativistic Fermi
surface calculated by Samokhin, Zijlstra and
Bose\cite{Samokhin2004a} shows three distinct sheets, each of which
becomes split into two as a result of the spin-orbit
interaction.
The spin-orbit coupling breaks the parity
symmetry $\hat{P}$, and therefore mixes singlet and
triplet pairing states.
A full
symmetry analyis\cite{Samokhin2004a,Samokhin2004b,Sergienko2004}
shows that  conventional
s-wave pairing ($\Delta_{\bf k}$ having full tetragonal symmetry)
and a p-wave triplet pairing state of the form
${\bf d}_{\bf k} \sim
(-k_y,k_x,0)$
are able to coexist.
If the microscopic pairing interaction
strongly favors one type of pairing over the other then
that one will be dominant, but in general both gap components are
present on a single Fermi surface sheet.

Following Hayashi {\it et al.}\cite{Hayashi2006} we take the pairing
state as the combined s-wave and ${\bf d}_{\bf k} \sim (-k_y,k_x,0)$
state described above. Using the notation of Ref.~\cite{Powell2003}
the mean-field BCS Hamiltonian for this system in the presence of a
magnetic Zeeman field can be written
\begin{widetext}
\begin{eqnarray}
H_{MF} & = &  \left(
\begin{array}{cccc}
\epsilon_{\bf k} +  H_z & -i \alpha g_{\bf k} e^{-i\phi} - i  H_y   &
i d_{\bf k} e^{-i\phi}
 & \Delta_{\bf k} \\
i\alpha g_{\bf k}e^{i\phi} + i  H_y & \epsilon_{\bf k}- H_z & -\Delta_{\bf k}
 & id_{\bf k} e^{i\phi} \\
-id^*_{\bf k} e^{i\phi} & -\Delta_{\bf k}^* & -\epsilon_{\bf k}-  H_z &
i\alpha g_{\bf k} e^{i\phi} +i  H_y\\
\Delta_{\bf k}^* & -i d^*_{\bf k} e^{-i\phi} & -i\alpha g_{\bf k}e^{-i\phi}- i  H_y &
-\epsilon_{\bf k}+  H_z \end{array} \right),
\label{eq:hnonzero}
\end{eqnarray}
\end{widetext}
where
for brevity we write ${\bf g}_{{\bf k},x} +i {\bf g}_{{\bf k},y} =i g_{\bf k} e^{i\phi} $, with
$\phi=\tan^{-1}{(k_y/k_x)}$. Similarly we write
 ${\bf d}_{{\bf k},x}+i{\bf d}_{{\bf k},y}=i d_{\bf k} e^{i\phi} $ and
we have set $H_x=0$.

In zero external field the diagonalization of this
Hamiltonian yields the quasiparticle energy dispersion,
\begin{equation}
E_{1,2}({\bf k}) = \sqrt{(\epsilon_{\bf k} \mp \alpha
 g_{\bf k})^2 + |  d_{\bf k}
\mp \Delta_{\bf k}|^2}. \label{eq:ekhzero}
\end{equation}
In the case where both singlet and triplet order parameters,
$\Delta_{\bf k}$ and $d_{\bf k}$,
can be chosen as real,  this is
the spectrum described by Hayashi {\it et al.}\cite{Hayashi2006}.
Assuming that the p-wave gap component
is dominant so that $\Delta_{\bf k}$ is smaller than the maximum
value of $d_{\bf k}$ on the Fermi surface, then one
of the Rashba split sheets of the Fermi surface
will have a line node, where
 $\Delta_{\bf k}=d_{\bf k}$
and the other sheet will be nodeless. For a spherical Fermi surface
and the simplest gap functions allowed by symmetry, $\Delta_{\bf
k}=\Delta$ and ${\bf d}_{\bf k}=d_0(-k_y,k_x,0)/k_F$, the nodes are
two horizonatal circles around the Fermi sphere at $\pm k_z = {\rm
const.}$ as illustrated in Fig. \ref{fig1}(a).
%We can now consider the topological structure of these line nodes.
As pointed out by Sato\cite{Sato2006} the nodes have $Z_2$
character. They can be removed by simply setting the relative phase
between the two complex scalar order parameters $\Delta_{\bf k}$ and
$d_{\bf k}$ non-zero, as one can readily see from
Eq.~\ref{eq:ekhzero}.

Let us now consider what will happen to the lines of nodes when time-reversal symmetry is broken by
applying a Zeeman exchange field ${\bf H}$. We assume that we are
dealing with relatively weak field and therefore the
Fulde-Ferrel-Larkin-Ovchinnikov state\cite{fulde,larkin,Kaur2005}
is not present.
 Let us first consider the case of a c-axis field. In this case  ${\bf H} \perp
{\bf g}_{\bf k}$ and the Hamiltonian (\ref{eq:hnonzero}) can be
diagonalized to obtain
\begin{widetext}
\begin{equation}
%\lefteqn{
E_{1,2}({\bf k}, H_z) =
%}
  \sqrt{\epsilon^2_{\bf k} + (\alpha k)^2 + H_z^2
+\Delta_{\bf k}^2 +k^2|d|^2 \mp \sqrt{4
H_z^2\left[\Delta_{\bf k}^2 +\epsilon^2_{\bf k}\right] +
\left[\Delta_{\bf k} k (d+d^*)+ 2\alpha k \epsilon_{\bf
k}\right]^2}}
\label{Ek:hz}
\end{equation}
\end{widetext}
and $E_{3,4}({\bf k}, H_z)=-E_{1,2}({\bf k}, H_z)$.
One can readily see that this spectrum has no node, and so the gap node
predicted by Hasyashi et al. is not topologically stable against
weak z-axis magnetic fields.

\begin{figure}[b]
\centerline{\scalebox{0.21}{\includegraphics{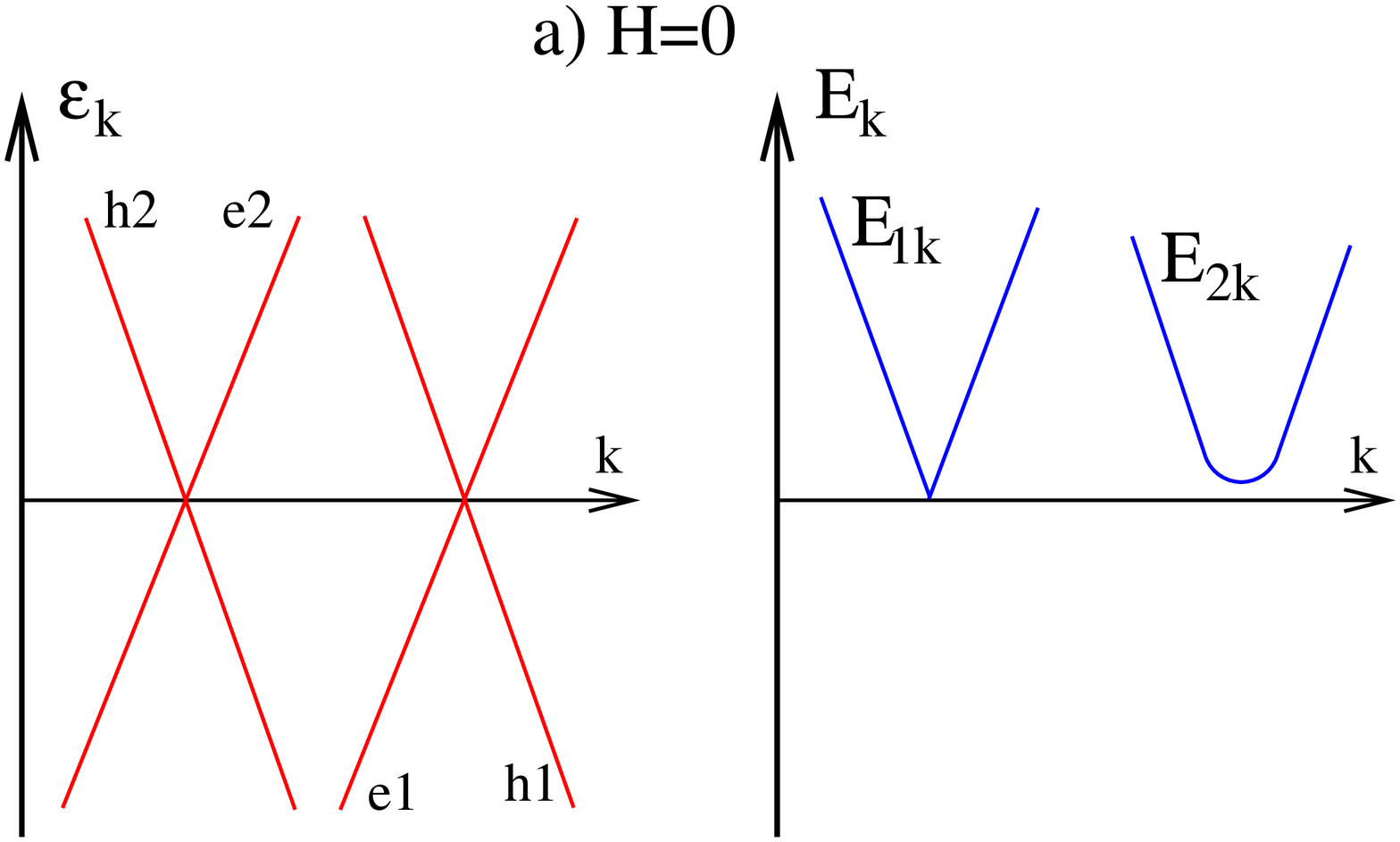}}}  %,width=8.5cm}
\centerline{\scalebox{0.21}{\includegraphics{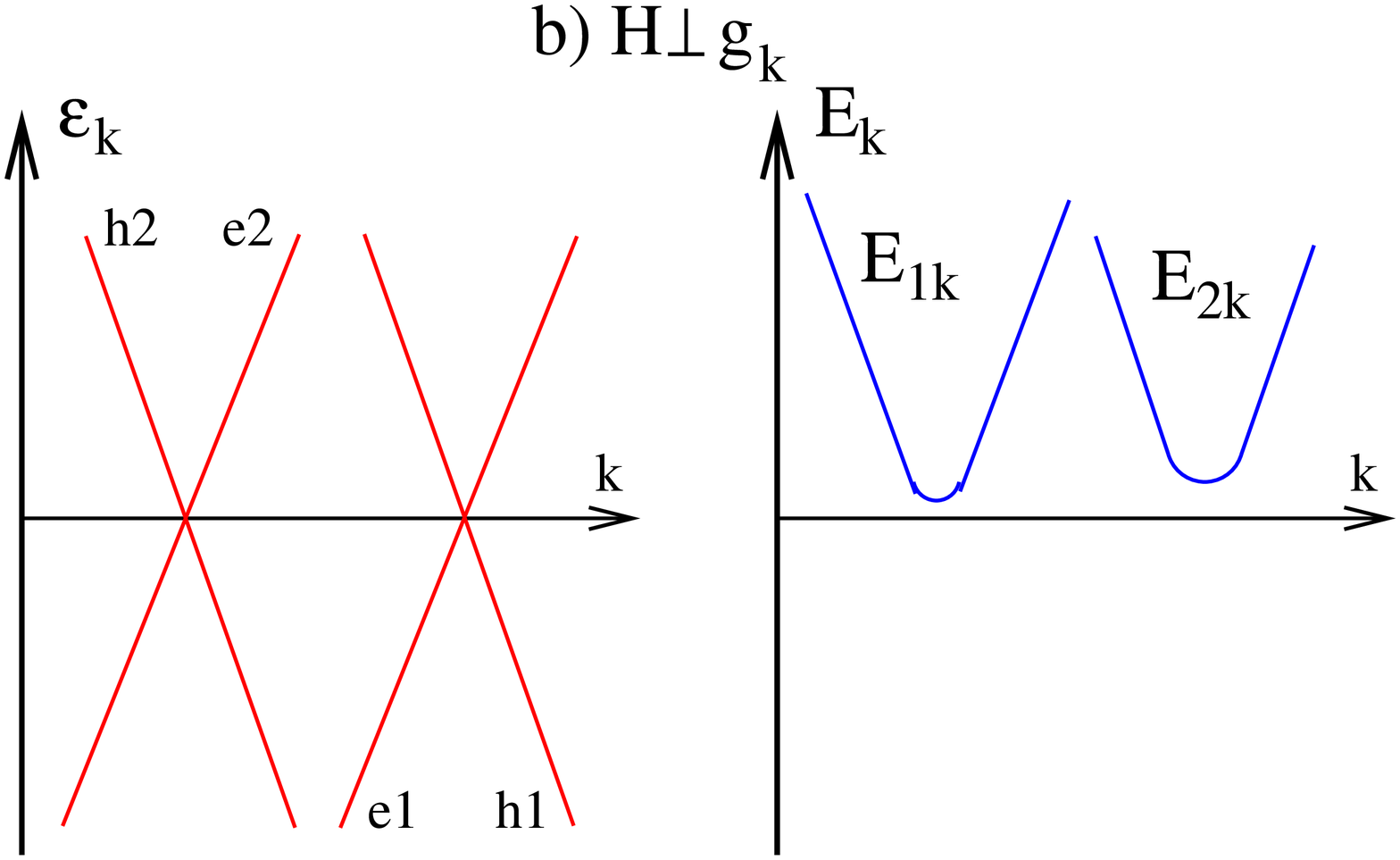}}}  %,width=8.5cm}
\centerline{\scalebox{0.21}{\includegraphics{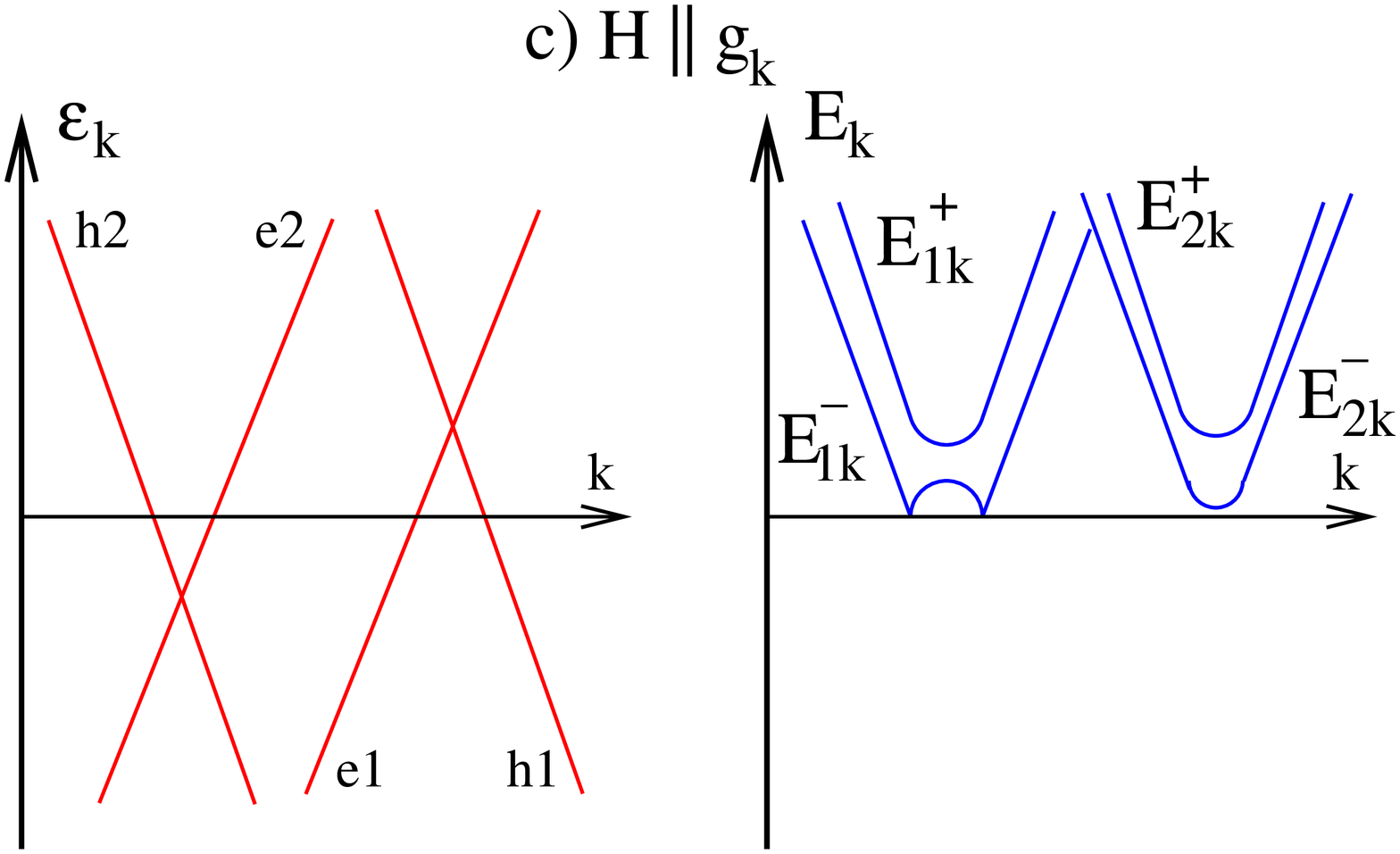}}}  %,width=8.5cm}
\caption{(color online) Normal state (left) and superconducting state (right)
quasiparticle dispersions, $\epsilon_k$ and $E_{\bf k}$, for
particles and holes as a function of $k=(k_x^2+k_y^2)^{1/2}$ for a
fixed $k_z$. From top to bottom, zero field ${\bf H}=0$, ${\bf H}
\perp {\bf g}_{\bf k}$, and ${\bf H} \parallel {\bf g}_{\bf k}$ }
\label{fig_ek}
\end{figure}

To explain the lack of topological stability, consider the
normal and superconducting state quasiparticle spectrum near the two
Rashba split sheets of the Fermi surface.  In
Fig.\ref{fig_ek}(a) where we show schematically the energy
dispersion E$_{\bf k_{||}}$ in a cut through the Fermi surface
at $k_z = \pm const$.
We have chosen the value of $k_z$ coinciding with
the line node predicted by Hayashi et al., given by the zero of Eq.~\ref{eq:ekhzero}.
 On the left hand panel of Fig.\ref{fig_ek}(a)  we show the normal state electron
(e1, e2) and hole (h1,h2) quasiparticle dispersions near the two Rashba Fermi surface
sheets.  On the right hand side we show the corresponding superconducting state
energies from Eq.~\ref{eq:ekhzero}. Clearly one sheet has a finite gap and the other has a node.
Now consider the spectrum when a weak c-axis field is applied, as shown in Fig.~\ref{fig_ek}(b).
In this case the spectrum becomes fully gapped, and the line node is removed. In effect the original
line node is ``accidental'', arising from a particular cancellation between the s-wave and p-wave gap
compoonents $\Delta_{\bf k}$ and $d_{\bf k}$ at some particular value of $k_z$.  When the symmetry breaking perturbation
$H_z$ is applied the crossing electron and hole-like levels $h2$ and $e2$ in Fig.~\ref{fig_ek}(a), interact
and so an avoided level crossing occurs, hence removing the line node.

On the other hand consider the case of a $a-b$ plane magnetic field.  The two Fermi surface
sheets will be perturbed differently depending upon whether ${\bf H} \parallel {\bf g}_{\bf k}$
or ${\bf H} \perp {\bf g}_{\bf k}$. The latter case is essentially identical to Fig.~\ref{fig_ek}(b), and so
for these parts of the Fermi surface the gap node is removed.  On the other hand, for the regions
where ${\bf H} \parallel {\bf g}_{\bf k}$ the gap node is found to \emph{bifurcate}. The reason
for this is illustrated in Fig.~\ref{fig_ek}(c). The field perturbs the two Rashba sheets oppositely, because they
have spin components parallel to the applied field ${\bf H}$. This
can be seen in the normal state spectrum to the left side of Fig.~\ref{fig_ek}(c). The corresponding
superconducting state spectrum  is shown to the right
in Fig.~\ref{fig_ek}(c).  The four quasiparticle states no longer obey the symmetry
$E_{3,4}({\bf k}, H_y)=-E_{1,2}({\bf k}, H_y)$. A result is that
one branch of the quasiparticle spectrum near the original line node now crosses zero twice, while the other
remains non-zero. This
implies that the gap node bifurcates.  Note that the final spectrum can always be defined as
positive, consistent with stability of the Fermi sea, after applying a suitable particle-hole transformation
to the quasiparticle states. This sign change results in the bifurcated node, as shown in the right panel in
Fig.~\ref{fig_ek}(c).

The final nodal topology for $a-b$ oriented magnetic fields is shown in
Fig.~\ref{fig1}(b). For ${\bf k}$ vectors with $x-y$ components parallel to ${\bf H}$ the gap node
is removed, while for ${\bf k}$ vectors with $x-y$ components perpendicular to ${\bf H}$ the gap node bifurcates.
Overall this leads to the unusual ``bomerang'' shaped gap line nodes as shown in Fig.~\ref{fig1}(b).
This topological structure is consistent with the general $Z_2$ topologigal arguments of Sato\cite{Sato2006}
and Volovik\cite{Volovik2006}. The $Z_2$ gap node has two topological invariants, which in the presence of
time reversal symmetry are independently conserved, leading to stability of the line node. But the
symmetry breaking perturbation ${\bf H}$ removes the stability, leading to either node removal or node bifurcation
depending on the field orientation.

\begin{figure}[t]
\centerline{\scalebox{0.46}{\includegraphics{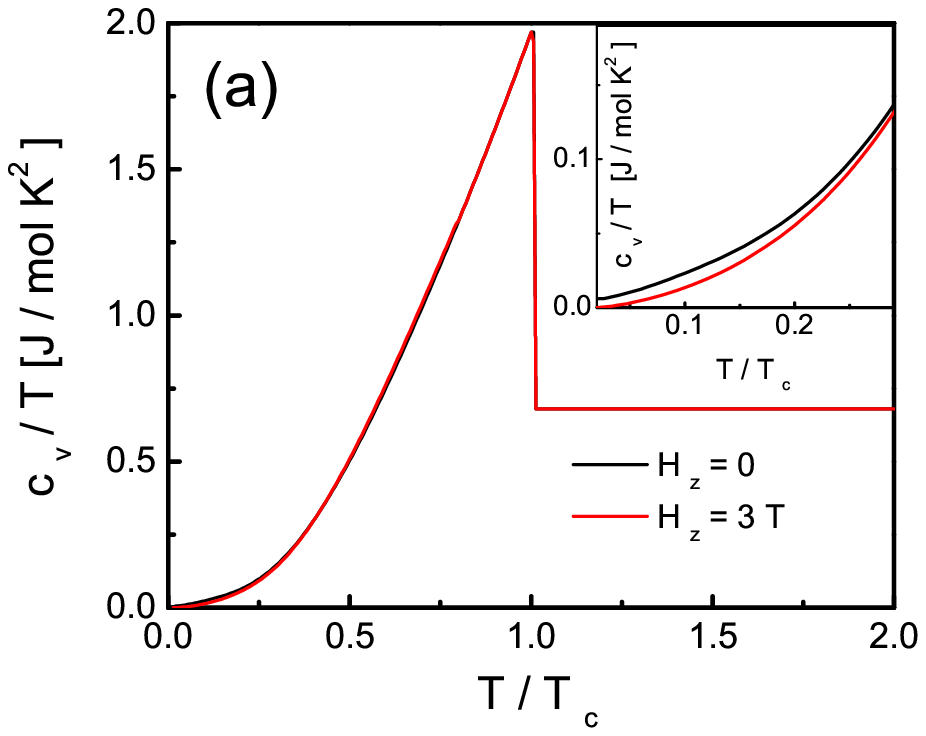}}\hspace*{-5mm}
\raisebox{-2mm}{\scalebox{0.48}{\includegraphics{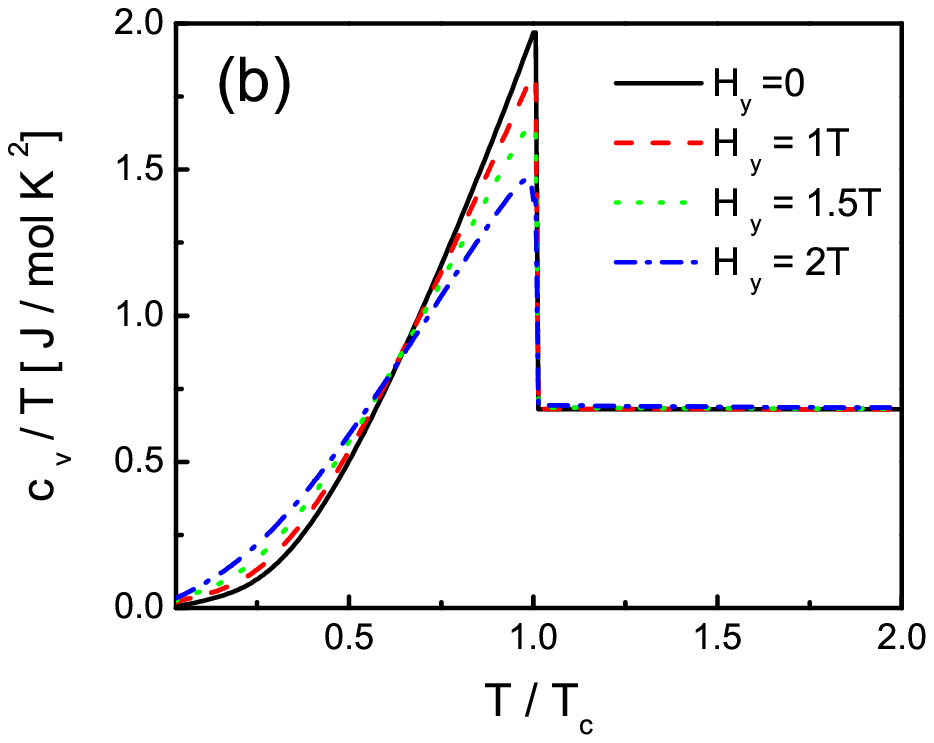}}}} %,width=12.5cm}
\caption{(color online) Calculated temperature dependence of the
specific heat with and without Zeeman field along $z$-direction(a)
and along $y$-direction(b). The inset shows the low-temperature
region for $H_z =0$ and $H_z = 3$T. Here, we assume the BCS-like
temperature dependence of the both $p$-wave and $s$-wave components
of the superconducting gap with ${\bf d}_{{\bf k}}(0) = 0.2$meV and
$\Delta_{{\bf k}}(0)=0.25\Delta_{p}(0)$ which gives about the BCS
ratio between the magnitude of the superconducting gap and $T_c =
0.75$K. Following experimental findings we adopt $E_f \approx
0.012$eV and estimate $k_F \approx 0.5 $ $\AA^{-1}$ which are
typical values for the heavy-fermion metals. We further approximate
the Rashba splitting of the bands $\alpha \sim 1 \times 10^{-3}$ eV
$\AA$.} \label{fig_sep}
\end{figure}

Experimentally this nodal change should be observable in unusual field dependence
of various low temperature properties. In CePt$_3$Si  NMR experiments show both a Hebel-Slichter
peak, a signature of the nodeless $s$-wave superconductor, and
power-law behavior of the spin-lattice relaxation at low
temperatures consistent with nodal structure of
the superconducting gap\cite{Yogi2004,Yogi2006}.
The predicted gap topology in Fig.~\ref{fig1} would imply a change in the
low temperature power law to exponential for $c$-axis fields, and a changed prefactor
of the low temperature power law for $a-b$ oriented fields.
Similarly the
changes in the line node topology upon applied magnetic field results in the
change of the thermodynamic characteristics such as heat capacity, $C_V / T$.
In Fig.\ref{fig_sep} we show our results for the temperature
dependence of the specific heat, $C_V / T$ for various Zeeman fields.
For $c$-axis fields the low temperature behavior changes from
linear to exponential, but the actual total change is small for physically
relevant values of ${\bf H}_z$, as shown in Fig.\ref{fig_sep}(a). In contrast
for $a-b$ plane fields the formation of the ``boomerang''-like
structure in the line nodes nodes both reduces the jump in the specific heat at $T_c$,
and leads to a noticeable increase in $C_V / T$  at low temperatures.
Furthermore, for larger
fields we also find that an even more substantial reduction of the jump occurs. This happens
because for large enough exchange fields a new pair of gap nodes develops
also on the second Rashba Fermi surface sheet. This is because for larger fields the lower
quasi-particle state $E_{2}^{-}({\bf k})$, also crosses zero as is evident from
Fig.\ref{fig_ek}(c). Most interestingly we find that these effects
develop at small enough fields in CePt$_3$Si to be observed
experimentally.

In conclusion, we have shown that Zeeman field induces an
interesting and novel phenomena in the the non-centrosymmetric heavy
fermion superconductor CePt$_3$Si. A c-axis field is
shown to remove the ``accidental'' line node, and change the
specific heat, $C_V/T$ from linear to exponential at low
temperatures. A field in the $a-b$ plane is shown
to spit the nodes into two 'boomerang'-like structures which will
 reduce the jump in the $C _ V /T$ and
enhance the value of $C _ V /T$ at zero temperature. These surprising changes
result from the changes of the $Z_2$ topological quantum numbers
by the time reversal symmetry breaking perturbation, consistent
with the arguments of Sato\cite{Sato2006} and Volovik \cite{Volovik2006}.

\acknowledgments JFA is grateful to the Max-Planck Institut f\"{u}r
Physik Komplexer Systeme, Dresden, for hospitality during the
preparation of this work. We would like to thank P. Fulde, G.E.
Volovik, and N. Perkins for helpful discussions.

\end{document}